\documentclass[fleqn,usenatbib]{mnras}

\usepackage{newtxtext,newtxmath}
\usepackage[T1]{fontenc}
\usepackage{float}
\DeclareRobustCommand{\VAN}[3]{#2}
\let\VANthebibliography\thebibliography
\def\thebibliography{\DeclareRobustCommand{\VAN}[3]{##3}\VANthebibliography}

\usepackage{graphicx}	
\usepackage{amsmath}	
\usepackage{amssymb}	

\title[Timing and evolution of PSR B0950+08]{Timing and the evolution of PSR B0950+08}

\author[Huang et al.]{
Hai-tao Huang$^{1, 2}$, Xia Zhou$^{1, 3, 4}$, \thanks{E-mail: zhouxia@xao.ac.cn(XZ)}
 Jian-ping Yuan$^{1, 3, 4}$, Xiao-Ping Zheng$^{5,6}$
\\
$^{1}$Xinjiang Astronomical Observatories, Chinese Academy of Sciences, Urumqi 830011, China\\
$^{2}$University of Chinese Academy of Sciences, 19A Yuquan Road, Beijing 100049, China\\
$^{3}$Key Laboratory of Radio Astronomy, Chinese Academy of Sciences, Nanjing 210008, China\\
$^{4}$Xinjiang Key Laboratory of Radio Astrophysics, Urumqi 830011, China\\
$^{5}$Department of Astronomy, School of physics, Huazhong University of Science and Technology, Wuhan 430074, China\\
$^{6}$Institute of Astrophysics, Central China Normal University, Wuhan 430079, China
}

\date{Accepted 2022 March 9. Received 2022 March 2; in original form 2021 December 31}

\pubyear{2015}

\begin{document}
\label{firstpage}
\pagerange{\pageref{firstpage}--\pageref{lastpage}}
\maketitle

\begin{abstract}
We present timing solutions of PSR B0950+08, using 14 years of observations at  Nanshan 26-m Radio Telescope of Xinjiang Astronomical Observatory. The braking index of PSR B0950+08 varies from --367 392 to 168 883, which shows an oscillation with large amplitude ($\sim 10^5$) and uncertainty. Considering the variation of braking indices and the most probable kinematic age of PSR B0950+08, a model withe long-term magnetic field decay modulated by short-term oscillations is proposed to explain the timing data. With this magnetic field decay model, we discuss the spin and thermal evolution of PSR B0950+08. The uncertainties of its age are also considered. The results show that three-component oscillations are the more reasonable for the spin-frequency derivative distributions of PSR B0950+08, and the initial spin period of PSR B0950+08 must be shorter than $97\rm\ ms$ when the age is equal to the lower bound of its kinematic age. The standard cooling model could explain the surface temperature of PSR B0950+08 with its most probable kinematic age. Vortex creep heating with a long-term magnetic field decay could maintain a relatively high temperature at the later stages of evolution and explain the thermal emission data of old and warm pulsars. Coupling with the long-term magnetic field decay, an explanation of the temperature of PSR B0950+08 with roto-chemical heating needs an implausibly short initial rotation period ($P_0 \lesssim  17\rm{ ms}$). The spin and thermal evolution of pulsars should be studied simultaneously. Future timing, ultraviolet or X-ray observations are essential for studying the evolution and interior properties of pulsars.
\end{abstract}


\section{Introduction}
\label{sec:introduction}

\par Pulsars are fast rotating magnetized neutron stars. To data, more than three thousand pulsars have been discovered (\url{http://www.atnf.csiro.au/research/pulsar/psrcat}, \citet{ATNF2005AJ}). An isolated pulsar loses its rotational energy through electromagnetic radiation, which can reduce the spin frequency of the pulsar \citep{Manchester1985,Lyne2015}. In general, the spin-down of a neutron star can be described by the torque equation
\begin{equation}
    \dot{\nu} = -K \nu^{n}
\end{equation}
where $\nu$ is the spin frequency of the pulsar, $K$ is a constant of proportionality related to the energy loss mechanisms of the pulsar, and $n$ is the braking index \citep{Lyne2015}. In the canonical case of a rotating magnetic dipole in a vacuum, the braking index is expected to be 3. Observationally, the braking index is derived from precise measurements of the spin frequency ($\nu$) and its time derivatives ($\dot{\nu}$, $\ddot{\nu}$), $n=\nu\ddot{\nu}/\dot{\nu}^2$ \citep{Blandford1988}. Precise measurements of  the spin frequency and its time derivatives ($\dot{\nu}$, and of the calculation of the braking index $n$ provide one of the best tests for theoretical spin evolution models \citep{Manchester1985,Lyne2015}. 
\par 
The research on the thermal evolution of neutron stars is an essential tool for understanding the fundamental properties of dense matter, because the cooling processes and heating mechanisms are usually determined by the composition and the equation of state of dense matter in its interior \citep{2004ARA&A..42..169Y,2006NuPhA.777..497P}. Some of the heating mechanisms, which are connected with rotational energy conversion \citep{1995ApJ...442..749R, Fernandez2005ApJ,2017ChPhC..41l5104Z}, are necessary to explain the thermal emission data of pulsars. We should simultaneously study the thermal evolution and long-term spin evolution of pulsars.

\par PSR B0950+08 (J0953+0755) is an isolated radio pulsar with period $P=253\rm\ ms$, period derivative $\dot{P}=2.3\times 10^{-16}\rm\ s/s$, characteristic age $\tau_{\rm c}=17.4 \rm\ Myr$, and a surface dipole magnetic field of $\sim 2.44\times10^{11}\rm\ G$  \citep{2004MNRAS.353.1311H}. An off-pulse emission search for PSR B0950+08 at 76 MHz shows that it has a surrounding pulsar wind nebulae (PWN) \citep{2020MNRAS.495.2125R}. Generally, an observable radio PWN
around such an old pulsar as PSR B0950+08 would not be expected, as most pulsars with radio PWN are $\sim 10^4 \rm yr$ old. The energy loss rate of the older PSR B0950+08, $\dot{E}= 5.6 \times 10^{32} \rm{erg ~s^{-1}}$, is coincident with its characteristic age found for younger pulsars with strong PWN \citep{2020MNRAS.495.2125R}. A previous study estimated that the kinematic age of PSR B0950+08 was much younger than the characteristic age \citep{Noutsos2013MNRAS}. Using
a Bayesian approach, \cite{2019MNRAS.482.3415I} found the most likely kinematic age $\tau_{\rm k}$ of PSR B0950+08 to be $1.9^{+5.5}_{-0.6}\ \rm{Myr}$ with $68$ per cent credible intervals. This value is significantly younger than the characteristic age $\tau_{\rm c}$ of B0950+08. The braking index of this pulsar computed through timing data was $n\approx -2333$ \citep{2004MNRAS.353.1311H}, but this value changed to $-1473.5$ in \citet{Hobbs2010MNRAS}. Because long timing observations would allow us to eliminate possible glitches and constrain the braking index, we must re-analyse more timing data to constrain the origin and spin evolution of PSR B0950+08.

\par The surface temperature of PSR B0950+08 has been given in the range of $(1- 3)\times 10^5 \rm\ K$, which is much higher than that predicted by the standard cooling model for a classical pulsar with characteristic age $17.4~\rm\ Myr$ \citep{2017ApJ...850...79P}. Recently, the best-fitting temperature of PSR B0950+08 has been updated to $(6-12)\times 10^4 \rm\ K$ \citep{Abramkin2022Apj}. These high temperatures of a pulsar older than $10^7 \rm yr$ mean that some heating mechanisms operate in interior of the neutron star, such as magnetic field decay \citep{2010A&A...522A..16G,2017ApJ...850...79P}, dark matter accretion \citep{Hamaguchi2019PhLB}, crust cracking \citep{cheng1992ApJ}, superfluid vortex creep \citep{1984ApJ...276..325A, Michelle1999,2010A&A...522A..16G}, deconfinement heating, and chemical heating \citep{1995ApJ...442..749R, Fernandez2005ApJ,wei2018MNRAS, Hamaguchi2019PhLB,wei2020, Yanagi2020MNRAS, Kantor2021MNRAS}. The kinematic age estimate in \citet{2019MNRAS.482.3415I} shows that the surface temperature of PSR B0950+08 can be explained within the standard cooling model and no additional heating mechanisms are required. Thus, we have to walk back through this pulsar's thermal and spin evolution.

\par Based on the timing observation, this paper aims to study the long-term spin and thermal evolution of PSR B0950+08 and try to explain evolution theoretically. The pulsar timing data are obtained from the Nanshan 26-m Radio Telescope of Xinjiang Astronomical Observatory (NSRT). We present the data reduction in Section \ref{sec:data}. Then, we give a plausible interpretation of braking indices of PSR B0950+08 and its long-term spin evolution in Section \ref{sec:rotation}. The thermal evolution of PSR B0950+08 is discussed in Section \ref{sec:thermal}. Conclusions and discussion are present in Section \ref{sec:conclusion}.

\section{data reduction}
\label{sec:data}
\par Timing observations of PSR B0950+08 were carried out by NSRT from 2000 January to 2014 January (from MJD 51547 to MJD 56664), with a dual-channel room-temperature receiver (2000 January to 2001 June) and a cryogenic receiver (2001 July to 2014 January). The monitoring of PSR B0950+08 was conducted with an average observing cadence of three times per month, and the duration of observations was 16-min in 18 per cent and 4-min in 82 per cent of cases. Early observations prior to 2010 were taken using an analogue filterbank (AFB) with 128 $\times$ 2.5-MHz subchannels and are described in \cite{Wangn2001}. After 2010 January 1, data were also obtained with a digital filterbank (DFB), which was configured to have 8-bit sampling and 1024 $\times$ 0.5-MHz channels, covering more than 320 MHz receiver bandwidth. The data are folded online with subintegration times of 1 min for the AFB and 30s for the DFB, then written to disc with 256 bins across the pulse profile for the AFB and 512 bins for the DFB \citep{2010ApJ...719L.111Y, 2017MNRAS.466.1234Y}.

\par The timing parameters are obtained from the full coherent timing analysis by TEMPO2 \citep{2006MNRAS.372.1549E,2006MNRAS.369..655H}. The timing parameters are shown in Table \ref{tab:timing results}. In order to show the evolution of frequency derivative $\dot{\nu}$ and braking index $n$, we divide the time-of-arrival (TOA) data into 30 segments, and the time span is about 150 d (see Table \ref{tab:brake}). We checked the TOA data, and no glitch was found during the whole time span.

\begin{table}
	\centering
    \caption{Timing Parameters for PSR B0950+08, in the Barycentric Coordinate Time (TCB) time-scale.}
    \label{tab:timing results}
    \begin{tabular}{ll}
    \hline
  Parameter  & Value \\
  \hline
  RAJ   & 09:53:09.30(2)       \\
  DECJ   &        +07:55:36.4(11)         \\
  PMRA ($\rm{mas\ yr^{-1}}$) & --2.0(8) \\
  PMDEC ($\rm{mas\ yr^{-1}}$)& 29.4(7) \\
  Data span (MJD)&51547  -   56664 \\
Frequency epoch (MJD) & 54438  \\
Frequency, $\nu$ ($\mathrm{Hz}$)          &   3.951548809(8)       \\
Frequency derivative, $\dot{\nu}\ (10^{-15}\mathrm{s}^{-2})$           &  --3.59(4)\\
Frequency second derivative, $\ddot{\nu}\ (10^{-26}\mathrm{s}^{-3}) $           & 1.0(6)   \\
Dispersion measure, DM ($\rm{cm^{-3}\ pc}$)      &       2.9692(8)\\
rms of timing residuals ($\rm{\mu s}$)& 2249.707\\
Braking index, n &  3339.8899(3)\\
Characteristic age, $\tau_{\rm c}\ (\mathrm{Myr})$ &17.4(2) \\
\hline
    \end{tabular}
\end{table}

\begin{table}
	\centering
	\caption{Timing parameters of PSR B0950+08 for 30 segments. Epoch was the centre date of each segment of data in Modified Julian Date; $\nu$, $\dot{\nu}$ and $\ddot{\nu}$ are the frequency of rotation and its first and second derivative respectively; the final column is the braking index $n=\nu\ddot{\nu}/\dot{\nu}^2$.}
\label{tab:brake}
\begin{tabular}{ccccc}
\hline
Epoch &$\nu(\rm{s}^{-1})$&$\dot{\nu}(10^{-15}\rm{s}^{-2})$ & $\ddot{\nu}(10^{-23}\mathrm{s}^{-3})$&$n\ (\times10^6$)\\ 
\hline
51869 & 3.951549607366(15) & $-$3.5906(7) &0.003(22) & 0.009(69)\\
52004 & 3.951549565481(10) & $-$3.5928(5) &$-$0.013(11) & $-$0.039(35)\\
52140 & 3.951549523245(8) & $-$3.5941(4) &$-$0.003(7) & $-$0.010(22)\\
52267 & 3.951549483823(7) & $-$3.5900(17) &$-$0.056(20) & $-$0.171(62)\\
52418 & 3.951549436926(4) & $-$3.5957(2) &$-$0.026(3) & $-$0.081(10)\\
52592 & 3.951549382831(5) & $-$3.5991(2) &$-$0.014(3) & $-$0.043(11)\\
52769 & 3.951549327772(6) & $-$3.5988(3) &0.011(4) & 0.033(14)\\
52941 & 3.951549274303(7) & $-$3.5975(3) &0.006(4) & 0.019(14)\\
53130 & 3.951549215558(6) & $-$3.5971(3) &0.003(4) & 0.009(13)\\
53265 & 3.951549173596(7) & $-$3.5952(2) &0.021(4) & 0.065(13)\\
53402 & 3.951549131046(7) & $-$3.5922(3) &0.028(5) & 0.084(15)\\
53550 & 3.951549085120(7) & $-$3.5883(4) &0.054(8) & 0.162(24)\\
53707 & 3.951549036502(8) & $-$3.5849(3) &0.011(10) & 0.034(31)\\
53887 & 3.951548980766(17) & $-$3.5820(13) &0.020(35) & 0.062(10)\\
54001 & 3.951548945506(13) & $-$3.5825(6) &$-$0.018(23) & $-$0.056(72)\\
54128 & 3.951548906164(10) & $-$3.5795(5) &0.055(16) & 0.169(49)\\
54271 & 3.951548861967(9) & $-$3.5833(3) &$-$0.042(9) & $-$0.129(29)\\
54372 & 3.951548830695(10) & $-$3.5876(4) & $-$0.066(14) & $-$0.204(44)\\
54505 & 3.951548789448(11) & $-$3.5976(5) & $-$0.120(15) & $-$0.367(48)\\
54625 & 3.951548752053(10) & $-$3.6039(4) & $-$0.027(11) & $-$0.083(35)\\
54755 & 3.951548711537(11) & $-$3.6065(4) & 0.011(17) & 0.032(51)\\
54922 & 3.951548659536(13) & $-$3.6027(4) & $-$0.006(19) & $-$0.017(59)\\
55016 & 3.951548630286(7) & $-$3.6026(4) & $-$0.024(10) & $-$0.073(32)\\
55170 & 3.951548582336(6) & $-$3.6041(3) & $-$0.016(7) & $-$0.049(22)\\
55312 & 3.951548538092(8) & $-$3.6036(3) & 0.027(8) & 0.082(27)\\
55458 & 3.951548492663(6) & $-$3.6015(3) & 0.0007(65) & 0.002(19)\\
55669 & 3.951548427005(5) & $-$3.6014(3) & 0.0001(47) & 0.0003(143)\\
55883 & 3.951548360422(5) & $-$3.5968(1) & 0.029(2) & 0.088(6)\\
56116 & 3.951548288069(5) & $-$3.5917(1) & 0.026(2) & 0.078(6)\\
\hline
\end{tabular}
\end{table}

\begin{figure}
\centering
\includegraphics[width=0.9\columnwidth]{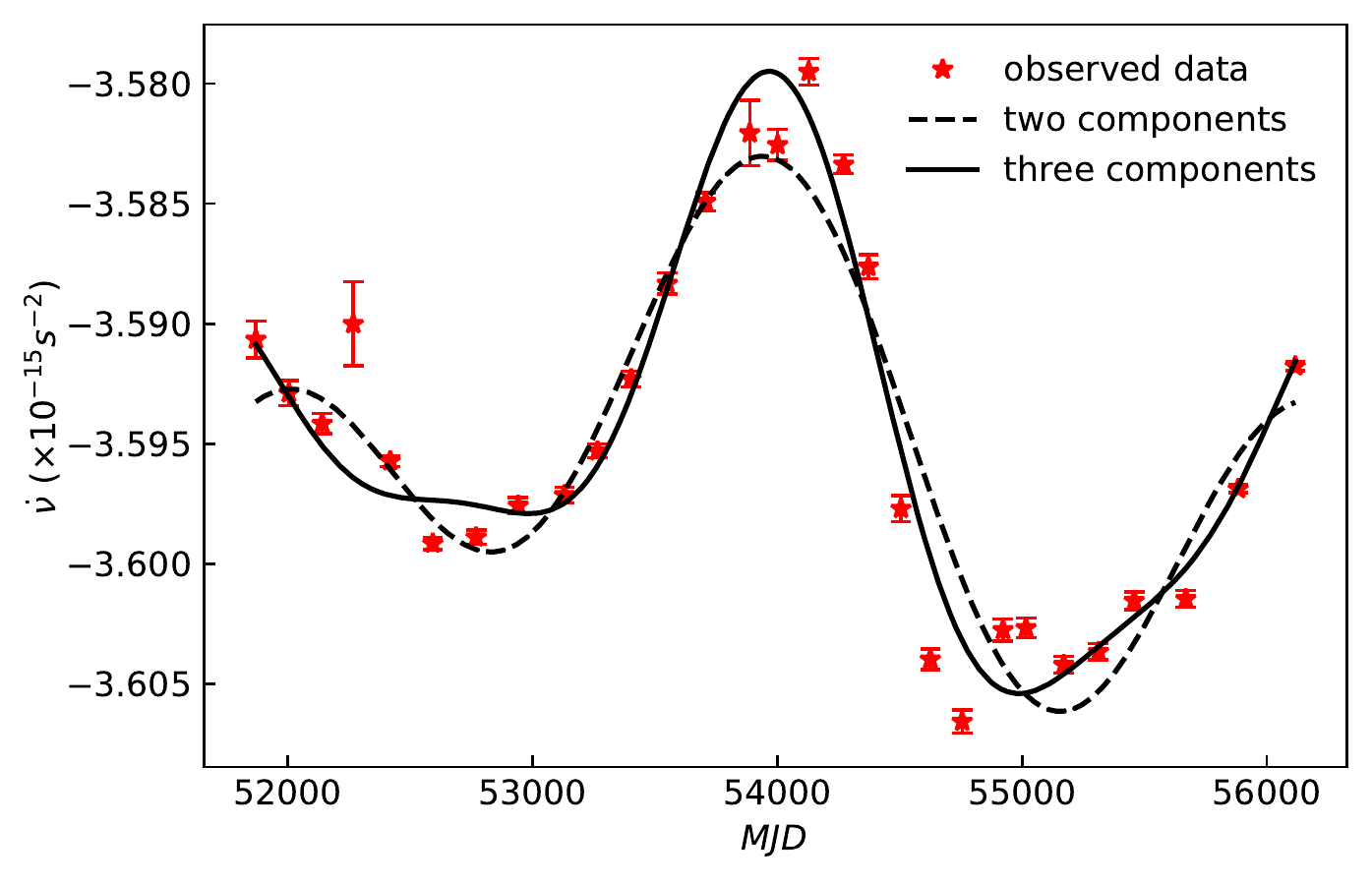}
\caption{Frequency derivative evolution of PSR B0950+08. The red asterisks are for the first-order frequency derivative obtained from NSRT timing data. The dashed and solid curves are for the two- and three-oscillation-component model, respectively.}
\label{fig:f1}
\end{figure}

\begin{figure}
	\includegraphics[width=\columnwidth]{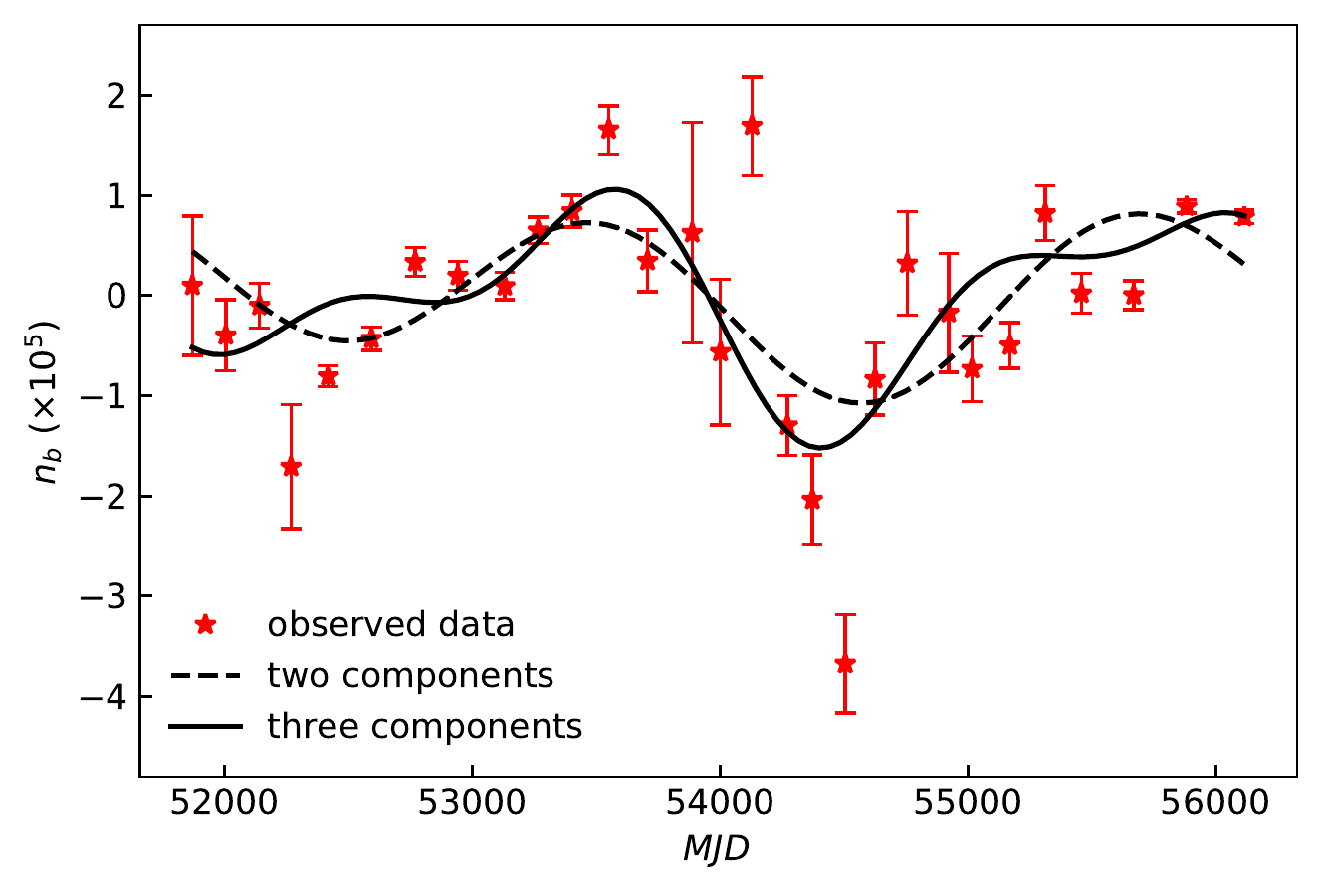}
    \caption{Braking index evolution of PSR B0950+08. The red asterisks are for the braking indices obtained from NSRT timing data. The dashed and solid curves are two- and three-oscillation-component model, respectively}
    \label{fig:nb}
\end{figure}

\section{Braking indices and long term spin evolution of PSR B0950+08}
\label{sec:rotation}
\subsection{The model for rotational evolution of PSR B0950+08}

\par The braking index of a pulsar usually remains constant for a very
long time. However, this is not true for PSR B0950+08, whose braking index varied from -367 392 to 168 883 during MJD51547 and MJD56664. Nearly half of the indices (14) are negative (see Table \ref{tab:brake}). The frequency and braking indices of PSR B0950+08 show a quasi-periodic oscillation, as shown in Fig. \ref{fig:f1} and Fig. \ref{fig:nb}. 
Former investigations suggested various hypotheses to explain the variation in the braking index: (1) magnetic field decay and oscillation \citep{2012ApJ...757..153Z,2013arXiv1312.3049X}, (2) evolution of the obliquity angle \citep{2020MNRAS.499.2826I}, (3) precession \citep{2020MNRAS.499.2826I}, (4) a complicated multipole structure of the poloidal magnetic field \citep{2020MNRAS.499.2826I}, and (5) fluctuations of neutron star magnetosphere \citep{Ou2016}. 
As suggested in \citet{2020MNRAS.499.2826I}, the evolution of
obliquity angle can cause large $n$ for certain initial obliquity angles
only in the non-physical case of vacuum magnetosphere. Precession in the plasma-filled magnetosphere \citep{2020MNRAS.499.2826I} and fluctuations of magnetosphere \citep{Ou2016} are considered, which could explain the large measured braking index. However, the precession also predicts an equal number of positive and negative braking indices. Meanwhile, these two mechanisms also change the pulsar profile and the evolution of inclination geometry. We checked the data, and no significant pulsar profile change is found, no glitch is detected in our data span from the timing data, and the measured braking indices are much larger than the precession predicted. Therefore, we do not think that changes in the magnetosphere can explain the measured braking indices of PSR 0950+08.

\par Moreover, the most probable kinematic age proposed in \citet{2019MNRAS.482.3415I} shows that PSR 0950+08 is younger than its characteristic age, and the most probable explanation for this situation is a combination of magnetic field decay and a long initial period. Therefore, we suggested that a phenomenological model of a long-term magnetic field decay modulated by short-term oscillations is the most plausible explanation for the oscillation of spin frequency-deviation and the significant variation of braking indices of PSR B0950+08.

According to the magnetohydrodynamic (MHD) simulations of pulsar magnetospheres \citep{2006ApJ...648L..51S}, the spin-down of pulsar is
\begin{equation}
\frac{d\nu}{dt}=-\frac{\pi^2 B(t)^2R^6\nu^3}{Ic^3}(\kappa_0+\kappa_1\sin^2\theta)=-AB(t)^2\nu^3
\label{eq:spinplasma}
\end{equation}
where $A=\pi^2R^6(\kappa_0+\kappa_1\sin^2\theta)/(Ic^3)$ is a constant, $\kappa_0$ and $\kappa_1$ are the numerical factors representing the effect of out-flowing particle: we take $\kappa_0\approx\kappa_1\approx1$ for simplicity \citep{2019ApJ...876..131K}. Then, the expression of braking index can be written as
\begin{equation}
n=3-\frac{2\dot{B}}{AB^3\nu^2}
\label{eq:brake}
\end{equation}
where $\dot{B}$ is the magnetic field time derivative.

\par An approximate description for long-term field evolution can be expressed as \citep{2008ApJ...673L.167A,2012MNRAS.421L.127P}
\begin{equation}
B_{\rm d}(t)=\frac{B_0\exp(-t/\tau_{\rm O})}{1+(\tau_{\rm O}/\tau_{\rm H})(1-\exp(-t/\tau_{\rm O}))}+B_{\rm fin}
\label{eq:exp}
\end{equation}
where $B_0$ is the initial field, $B_{\rm fin}$ is relic field and we set as $3.4355\times10^{11}\ \rm{G}$ in this work, 
$\tau_{\rm O}$ is the Ohmic decay time-scale and $\tau_{\rm H}$ is the Hall drift times-cale. The short-term oscillation of magnetic field can be written as \citep{1996ApJ...473..322T,2004ApJ...609..999C,2013arXiv1312.3049X}
\begin{equation}
B(t)=B_{\rm d}(t)\Big[1+\sum_{ i} k_{ i}\cos\big(\omega_{ i} t+\phi_{ i}\big)\Big]
\label{eq:oscillation}
\end{equation}
where $k_{ i}$, $\omega_{ i}$ and $\phi_{ i}$ is the amplitude, frequency and initial phase of the $i$-th oscillating component respectively, and the second term of equation~\ref{eq:oscillation} might be composed of multiple oscillation components.

\cite{2004ApJ...609..999C} suggested that the perturbation of the background magnetic field will induce the Hall wave in the crust of neutron star, and the presence of different modes of Hall wave determined by the given boundary condition (the number of nodes $m$). The period of oscillation is $P_{\rm{H}}\simeq\tau_{\rm{H,b}}/m^2\simeq 10^7\ \mathrm{yr}/(B_{12}m^2)$, the amplitude of oscillations are $\delta B\propto m^{-7/6}$ and $\delta B\propto m^{-1}$ for strong and weak cascade condition respectively, and the maximum value is $\delta B\sim 10^{12}\rm{G}$ \citep{2004ApJ...609..999C}. It has also been suggested that the oscillations of the magnetic field from the Hall wave are probably responsible for the quasi-periodic oscillations of spin-frequency derivative in observation \citep{2013arXiv1312.3049X,Xie2014AN}. For PSR B0950+08, we choose the number of node about $m\sim2000$, with a period about $7~\rm yr$, and the amplitude of oscillation $k\sim\delta B/B\sim m^{-1}\sim 10^{-4}-10^{-3}$. The oscillation term in equation \ref{eq:oscillation} can be ignored in the long-term spin evolution because the amplitude is small in our work.

\subsection{Modeling the $\dot{\nu}$ and braking indices of PSR B0950+08 }

The distributions of frequency derivative and braking indices are shown in Fig.s \ref{fig:f1} and \ref{fig:nb}. We have also plotted the evolution curves of the frequency derivative and braking indices from the model introduced in the above section. We found that the spin-frequency derivative evolution depends only on the number of oscillation components and their parameters. \citet{2016MNRAS.459..402G} suggested that the number of main peaks in the power spectrum equals the number of oscillation components in the short-term-oscillation magnetic field model. Because of the lack of the power spectrum of PSR B0950+08, two- and three-oscillation-component magnetic field models are considered while fitting the distribution of $\dot{\nu}$. The best-fitting parameters are presented in Table \ref{tab:osci}, and the $R$-squared of each model are also given. The fitting results show that three-component oscillations are more reasonable for the spin-frequency derivative distribution of PSR B0950+08. In Fig. \ref{fig:nb}, there is a sizeable relative uncertainty with the braking indices, and we found that these uncertainties come from the uncertainty of $\ddot{\nu}$. 

\begin{table}
\centering
\caption{Best fitting parameters for the short-term magnetic field oscillation.}
\begin{tabular}{ccc}
\hline
Parameters & Two-components &  Three-components \\
\hline
$k_1$ & $9.976\times10^{-4}$ & $1.19\times10^{-3}$ \\
$k_2$ & $8.36394\times10^{-4}$ & $7.04821\times10^{-4}$ \\
$k_3$ & ---- & $3.78362\times10^{-4}$ \\
$\omega_1$  & $3\times10^{-3}$  & $2.83\times10^{-3}$ \\
$\omega_2$  & $1.48\times10^{-3}$  & $1.47\times10^{-3}$ \\
$\omega_3$  & ---- & $4.97\times10^{-3}$ \\
$\phi_1$ & $-12.9376$ & $2.77473$ \\
$\phi_2$ & $-5.23759$ & $8.00324$ \\
$\phi_3$ & ---- & $18.97725$ \\
\hline 
$\textit{R}^2$ &0.8663 &0.93581 \\
\hline
\label{tab:osci}
\end{tabular}
\end{table}

\par We also gave out the braking index cover all the 14 years data span, which is $n\sim3.3\times10^3$. The distribution of brake index also shows a significant negative at about $\rm MJD~54500$, but there was no out line value of $\dot{\nu}$ at the same time, so we suggest that the irregular may cause the significant negative value of braking indices. The braking indices are dominated by the second derivative of spin frequency $\ddot{\nu}$, because it shows an oscillation around zero and the $\ddot{\nu}$ may contain different kinds of timing noise, which will cause the irregular distribution of braking index.

\subsection{Long-term rotational evolution of PSR B0950+08}

\par For the long-term rotational evolution, we assume PSR B0950+08 to be a typical NS with mass $M=1.4M_{\sun}$, radius $R=10\rm\ km$, and inclination angle $\theta=\pi/4$. Combining equation\ref{eq:spinplasma}, \ref{eq:exp} and \ref{eq:oscillation}, we obtain
 \begin{equation}
 \begin{split}
 \frac{1}{\nu^2(t)}-\frac{1}{\nu^2_0}=&2AB^2_0\tau_H\Bigg\{\frac{(1+\zeta)(1-\eta)}{1+\zeta(1-\eta)}-\frac{1}{\zeta}\ln\big[1+\zeta(1-\eta)\big]\Bigg\}\\&+2AB^2_{\rm fin}t+4AB_0B_{\rm fin}\tau_H\ln\big[1+\zeta(1-\eta)\big]
 \end{split}
 \label{eq:nuexp}
 \end{equation}
where $\zeta=\tau_O/\tau_H$. As discussed in \citet{2004ApJ...609..999C}, \citet{ 2015AN....336..831I} and \citet{2020MNRAS.499.2826I}, $\zeta=\tau_O/\tau_H \sim (10^{-4}-10^{-3}) \ll1$ and $\tau_O$ varies over a wide range. Considering the age of this pulsar $t \gtrsim 1\rm Myr$, $\eta=\exp(-t/\tau_O)\ll1$ is small quantity. Then, equation \ref{eq:nuexp} can be written as 
 \begin{equation}
 \frac{1}{\nu^2(t)}-\frac{1}{\nu^2_0}\simeq2A\Big[\frac{1}{2}B^2_0\tau_O+B^2_{\rm fin}t+2B_0B_{\rm fin}\tau_O\Big]
 \label{eq:nuexpapp}
 \end{equation}
and then the true age $t_{\rm true}$ from the evolution model can be written as
 \begin{eqnarray}
 t_{\rm true}&=&\frac{P^2-P^2_0}{2AB^2_{\rm fin}}-\frac{\tau_O}{2B^2_{\rm fin}}(B^2_0+4B_0B_{\rm fin}) \nonumber \\ 
 &=&\tau_{\rm c}-\frac{P_{0}}{2AB^{2}_{\rm fin}}-\frac{\tau_{\rm O}(B^2_0+4B_0B_{\rm fin})}{2B^2_{\rm fin}}\label{eq:pre}
 \end{eqnarray}
where $P_0$ and $B_0$ are the initial spin period and magnetic field, respectively. The first term in the right-hand side of equation \ref{eq:pre} is the characteristic age, the second term is the effect of the initial period, and the third term shows the speed up of braking due to the magnetic field decay.

Different spin evolution parameters, such as the initial magnetic field $B_{\rm 0}$, the decay timescale $\tau_{\rm O}$, and the initial spin period $P_{\rm 0}$, will result in different evolution tracks and the true age $t_{\rm true}$ from the evolution model. The most probable kinematic age of B0950+08 is $\sim 2\rm Myr$, and the credible interval that contains 68 per cent of the probability density is [1.2$\rm Myr$, 8.0$\rm Myr$], and the kinematic age with the credible interval that contains 95 per cent of the probability is [0.37$\rm Myr$, 17.0$\rm Myr$] \citep{2019MNRAS.482.3415I}. The upper bound of ages with 95 per cent of the probability is closed to the characteristic one. Usually, an additional braking mechanism will speed up the braking process and make the stars look younger than their characteristic age. Considering the possible kinematic and characteristic age of PSR B0950+08, a relatively low magnetic field decay time-scale is chosen. 

\begin{figure}
   \centering
   \includegraphics[width=\columnwidth]{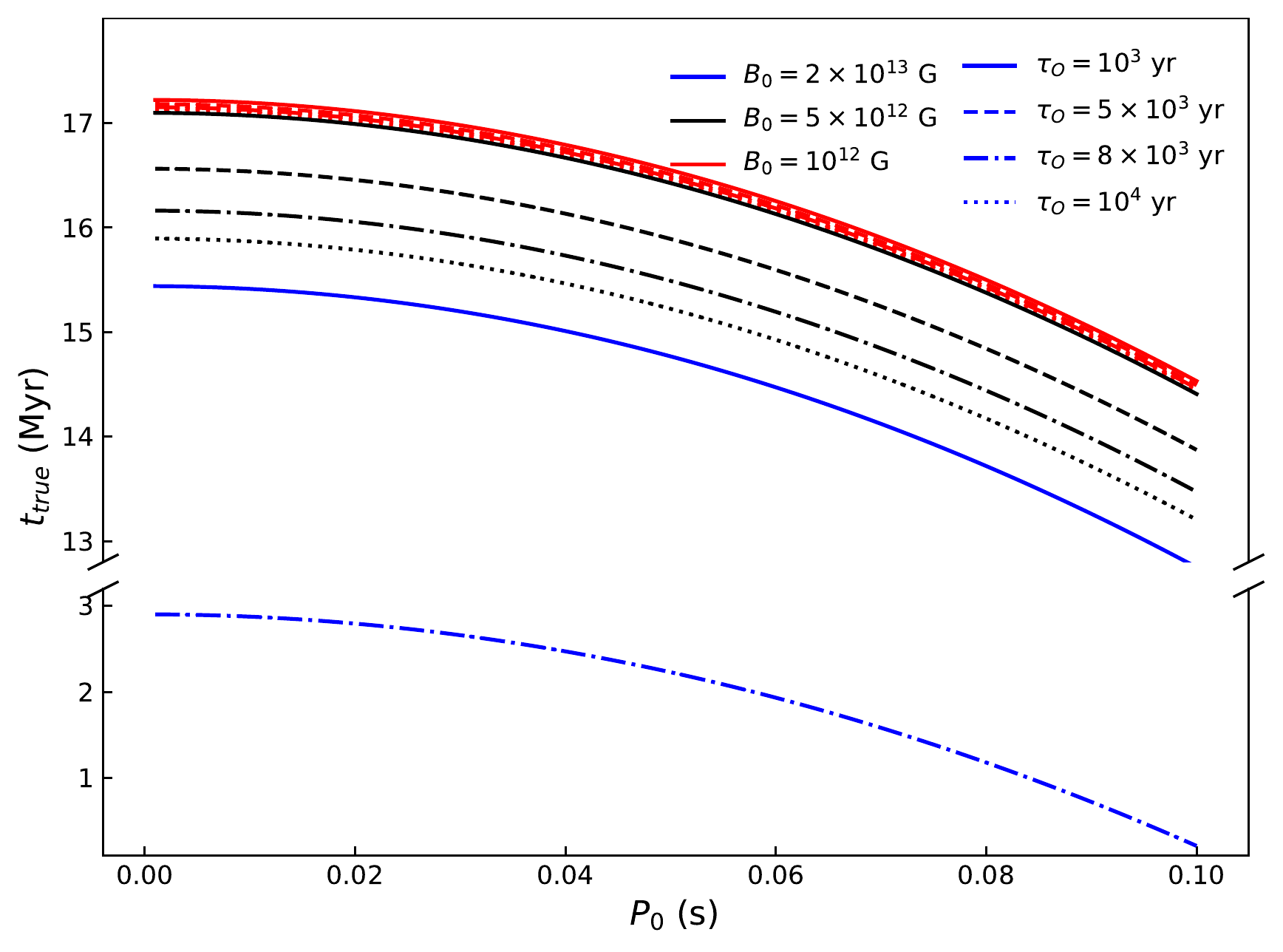}
    \caption{The true age $t_{\rm true}$ from the magnetic field decay model versus initial period. The red, black and blue lines are for different initial magnetic field $B_0=10^{12}\rm\ G$, $5\times10^{12}\rm\ G$ and $2\times10^{13}\rm\ G$, respectively. The solid, dashed, dash-dotted and dotted lines are for different Ohmic decay time-scale $\tau_{\rm O}=10^3\rm\ yr$, $5\times10^3\rm\ yr$, $8\times10^3\rm\ yr$ and $10^4\rm\ yr$.}
    \label{fig:tpre_exp}
\end{figure}

From equation \ref{eq:pre} and Fig. \ref{fig:tpre_exp}, we find $t_{\rm true}\sim (0.21-15.43)\ \rm Myr$ while $\tau_{\rm O}$ varies from $10^3\rm{yr}$ to $10^4\rm{yr}$ with $B_0=2\times10^{13}\rm\ G$, and $t_{\rm true}$ is affected mainly by the Ohmic decay time-scale. The true age for the evolutionary model is $t_{\rm true}\sim(13.47-17.10)\rm\ Myr$, while $\tau_{\rm O}$ varies from $10^3\rm{yr}$ to $10^4\rm{yr}$ with $B_0=5\times10^{12}\rm\ G$, and $t_{\rm true}$ is affected mainly by $P_0$ and $\tau_O$. With $B_0=10^{12}\rm G$, $t_{\rm true}\sim (14.44-17.22)\rm Myr$ which is affected mainly by $P_0$. With the lower limit of kinematic age, the initial spin period $P_0$ must smaller than $97\rm\ ms$ when $B_0=2\times10^{13}\rm\ G$ and $\tau_O=8\times10^3\rm\ yr$. 

\begin{figure*}
\includegraphics[width=190mm]{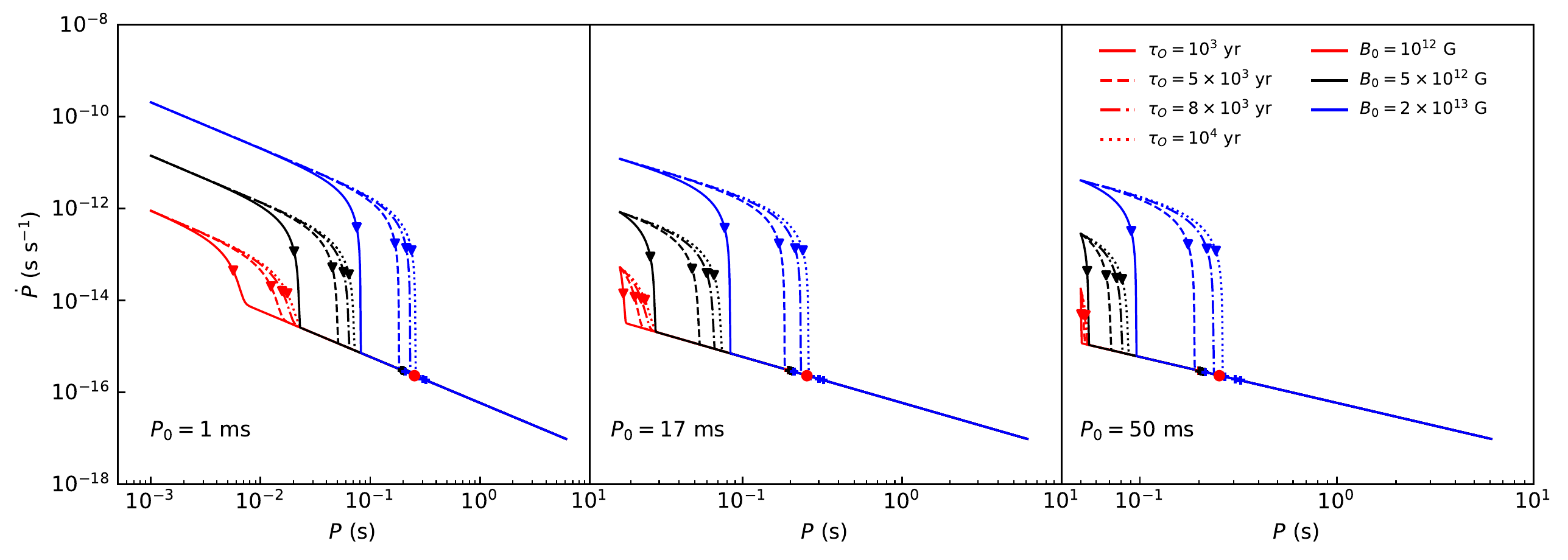}
\centering
\caption{Spin evolution curves of PSR B0950+08 with different initial spin period in a $P-\dot{P}$ diagram. The red, black and blue lines are for different initial magnetic field ,namely $B_0=10^{12}\rm\ G$, $5\times10^{12}\rm\ G$ and $2\times10^{13}\rm\ G$. The solid, dashed, dash-dotted and dotted lines are for different Ohmic decay time-scales, $\tau_{\rm O}=10^3\rm\ yr$, $5\times10^3\rm\ yr$, $8\times10^3\rm\ yr$ and $10^4\rm\ yr$. The triangles and crosses represent its periods and period derivatives at ages of corresponding Ohmic time-scale and $10^6\rm\ yr$, respectively. The red dot is PSR B0950+08.}
\label{fig:pdot_para}
\end{figure*}

In Fig. \ref{fig:pdot_para}, we give out the spin evolution curves of PSR B0950+08 with different initial spin period in a $P-\dot{P}$ diagram. It shows that different initial spin periods significantly affect the early spin evolution, but the evolutionary trend stays almost the same. Within the magnetic field decay model, the rapid spin-down ends around $\tau_O$ and then maintains an iso-magnetic evolution. 

\section{Thermal Evolution of PSR B0950+08}
\label{sec:thermal}
\par The age of PSR B0950+08 is still in debate. To discuss the thermal evolution of PSR B0950+08, we consider two cases: a younger pulsar without a heating mechanism, and an older but warm pulsar with various heating mechanisms. The heating mechanisms for the old neutron star are usually connected with its spin evolution. In this section, we consider the spin and thermal evolution of pulsars simultaneously, and the spin-down of the pulsar is based on the model discussed in the previous section. 

The thermal evolution equation is \citep{2004ARA&A..42..169Y},
\begin{equation}
\label{eq:thermal}
C_V\frac{\mathrm{d}T}{\mathrm{d}t}=-L_{\nu}-L_{\gamma}+H
\end{equation}
where $C_V$ is the total heat capacity of the star, $L_{\nu}$ and $L_{\gamma}$ are the luminosity of neutrino and photon emission respectively, and $H$ is the power generated by internal heating mechanisms, of which we consider the following two.

\begin{itemize}
    \item[(i)] \textit{Chemical heating}. Chemical heating is a heating mechanism that is associated with reactions produced by density changes \citep{1995ApJ...442..749R,Fernandez2005ApJ}. When neutron stars are spinning down, their internal density will increase. The density variation changes the chemical equilibrium state throughout the core. If the finite departure from the chemical equilibrium which modifies the reaction is large enough, the net effect of the reactions will increase the chemical energy at the expense of the stored rotational energy. This is a channel through which neutron star’s rotational energy can be converted into heat \citep{1995ApJ...442..749R,Fernandez2005ApJ,2017ChPhC..41l5104Z}. In order to solve the thermal balance equation \ref{eq:thermal} with the chemical heating mechanism, we use the formulas in \citep{1995ApJ...442..749R}
\begin{equation}
H_{\rm ch} = -\delta\mu\left(\alpha n \frac{E_{nx}}{E_{xx}}\frac{\Omega\dot{\Omega}}{G\rho_{\rm c}}+\frac{\Gamma}{n}\right)
\label{eq:ch}
\end{equation}
where the first term in the parentheses accounts for the change in the chemical equilibrium due to spin-down, and the second term accounts for the change in the actual chemical state due to different reaction processes. A detail description of the model can be found in \citet{1995ApJ...442..749R}.

\item[(ii)]\textit{Vortex creep heating}.
  The presence of a neutron superfluid in the interior of neutron stars is generally accepted, and the vortex creep model is widely used to explain pulsar glitches \citep{1984ApJ...276..325A,2006MNRAS.372..489A}. It is suggested that vortices are not pinned to the nuclear lattice all the time, but pin and unpin. While the star is spinning down, the vortex line moves out and through the inner crust, and they will unpin when the difference between the angular velocity of superfluid part $\Omega_{\rm s}$ and the angular velocity of the crust $\Omega_{\rm c}$ reaches a critical value. The interaction of vortex lines of the rapid spinning superfluid matter with the more slowly spinning normal matter in the inner crust of neutron stars can heat the star. The heating luminosity caused by the vortices pinning or unpinning to the nuclear lattice is given by \citep{1984ApJ...276..325A,2010A&A...522A..16G}
  \begin{equation}
    H_{\rm vc}= J|\dot{\Omega}|
\label{eq:vc}
    \end{equation}
    where $J\simeq \bar{\omega}I_{\rm p}$ with $I_{\rm p}$ the moment of inertia of the pinning
layer. For simplicity, we use a numerical formula to describe the heating effect of vortex creep in this work, namely $H_{\rm vc}\simeq\Big(10^{29}- 10^{31}\Big)\big|\dot{\Omega}_{-14}\big|\  \mathrm{erg\ s^{-1}}$ \citep{2010A&A...522A..16G}, where $\dot{\Omega}_{-14}$ is the angular velocity derivative in units of $10^{-14}\rm\ rad\ s^{-2}$. 
\end{itemize} 

\begin{figure}
	\includegraphics[width=\columnwidth]{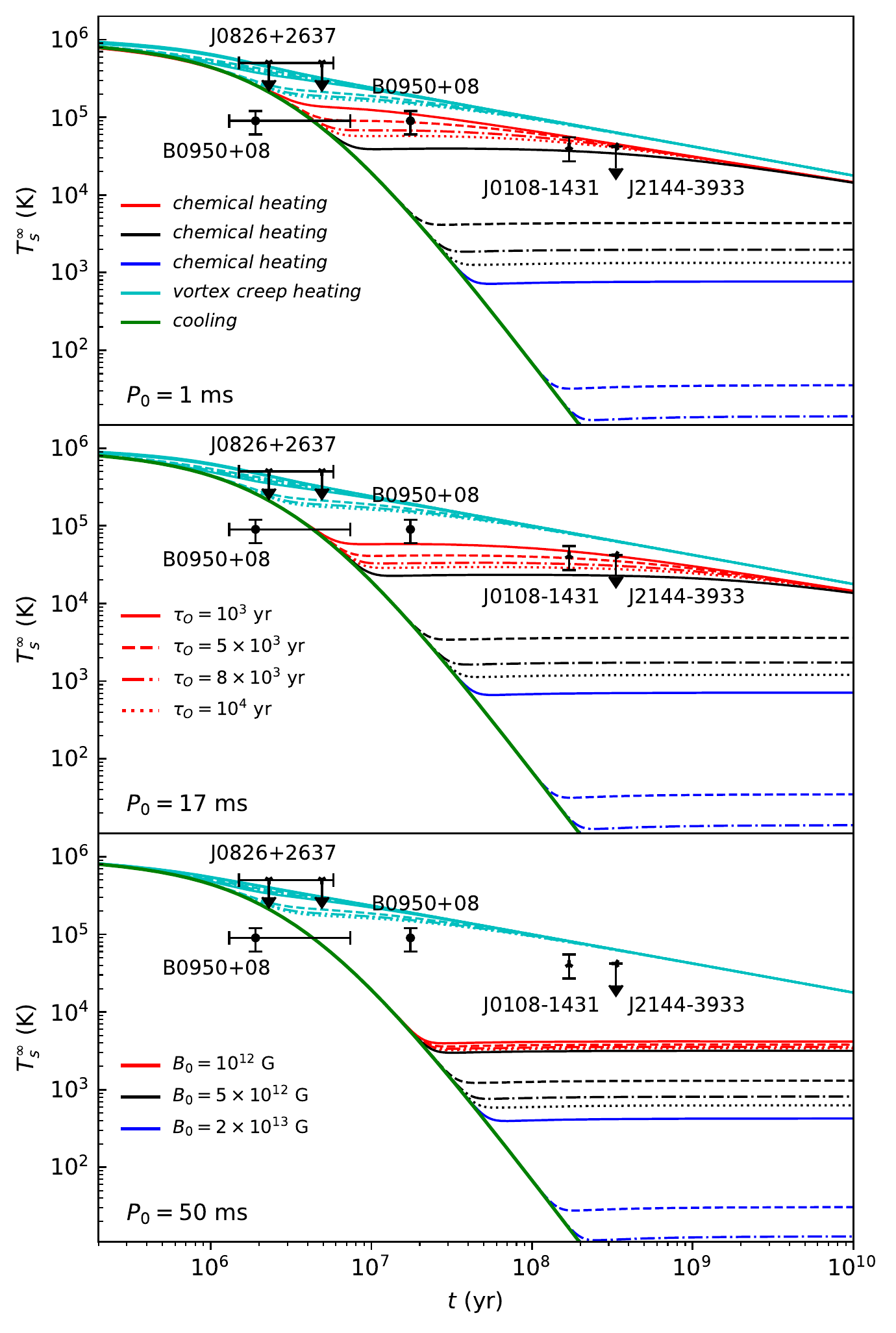}
    \centering
    \caption{Thermal evolution curves of neutron stars with different spin evolution parameters. The green lines are for the standard cooling without any heating, the cyan lines are for the case with vortex creep heating, and the others are for roto-chemical heating. The red, black and blue lines are for the different value of initial magnetic field, $B=10^{12}\rm\ G$, $5\times10^{12}\rm\ G$ and $2\times10^{13}\rm\ G$, respectively. The solid, dashed, dash-dotted and dotted lines are  for the different values of Ohmic decay time-scale, namely $10^3\rm\ yr$, $5\times10^3\rm\ yr$, $8\times10^3\rm\ yr$ and $10^4\rm\ yr$, respectively. Several old but still warm pulsars are also plotted, and the error bars show the uncertainties of their age and surface temperature. The surface temperatures with kinematic age and characteristic age of PSR J0953+0755 and PSR J0826+2537 are also plotted.}
    \label{fig:Ts_evolve_total}
\end{figure}

The thermal evolution curves with or without different heating mechanisms are shown in Fig. \ref{fig:Ts_evolve_total}, where $T_s^{\infty}=T_s\sqrt{1-2GM/(Rc^2)}$ is the surface temperature at infinity. The relations between the surface temperature and the internal temperature is given by $T_{8}=1.288(T^4_{\rm s,6}/g_{\rm s,14})^{0.455}$ \citep{1983ApJ...272..286G}, where $g_{\rm s,14}$ is the proper surface gravity of the star in units of $10^{14}~\rm cm~s^{-2}$ \citep{Potekhin2001AA}, and $T_8$ and $T_{\rm s,6}$ are the internal and surface temperature of the star in units of $10^8 \rm K$ and $10^6 \rm K$. The observed surface temperature with characteristic age and kinematic ages with its 68 per cent credible interval are also plot in Fig. \ref{fig:Ts_evolve_total}. In order to compare the effect of different heating mechanisms, we also plotted some other old but still warm pulsars in Fig. \ref{fig:Ts_evolve_total}, including PSR J0826+2637 \citep{2004ApJ...615..908B}, PSR J0108--1431 \citep{2021ApJ...911....1A}, PSR J2144--3933 \citep{2019ApJ...874..175G} and PSR J0437--4715 \citep{Kargaltsev2004,Durant2012ApJ,Guillot2016MNRAS}. 

As shown in Fig. \ref{fig:Ts_evolve_total}, the surface temperature of PSR B0950+08 can always be explained by the standard cooling model with the long-term magnetic field decay, while the age of PSR B0950+08 is around its most probable kinematic age. In this case, no heating mechanism is needed. We also find that a model of vortex creep heating with a long-term magnetic field decay could explain the surface temperature of some other old but warm pulsars and the surface temperature of PSR B0950+08 while its age is a characteristic one. With the heating effect of the vortex creep heating, the surface temperature would gradually decrease from $8.5\times10^5\rm\ K$ to $2.3\times10^4\rm\ K$ during $10^6-10^{10}\rm\ yr$. The effect of different initial spin periods and magnetic field decay parameters on the vortex creep heating is insignificant. 

\begin{figure}
\centering
\includegraphics[width=\columnwidth]{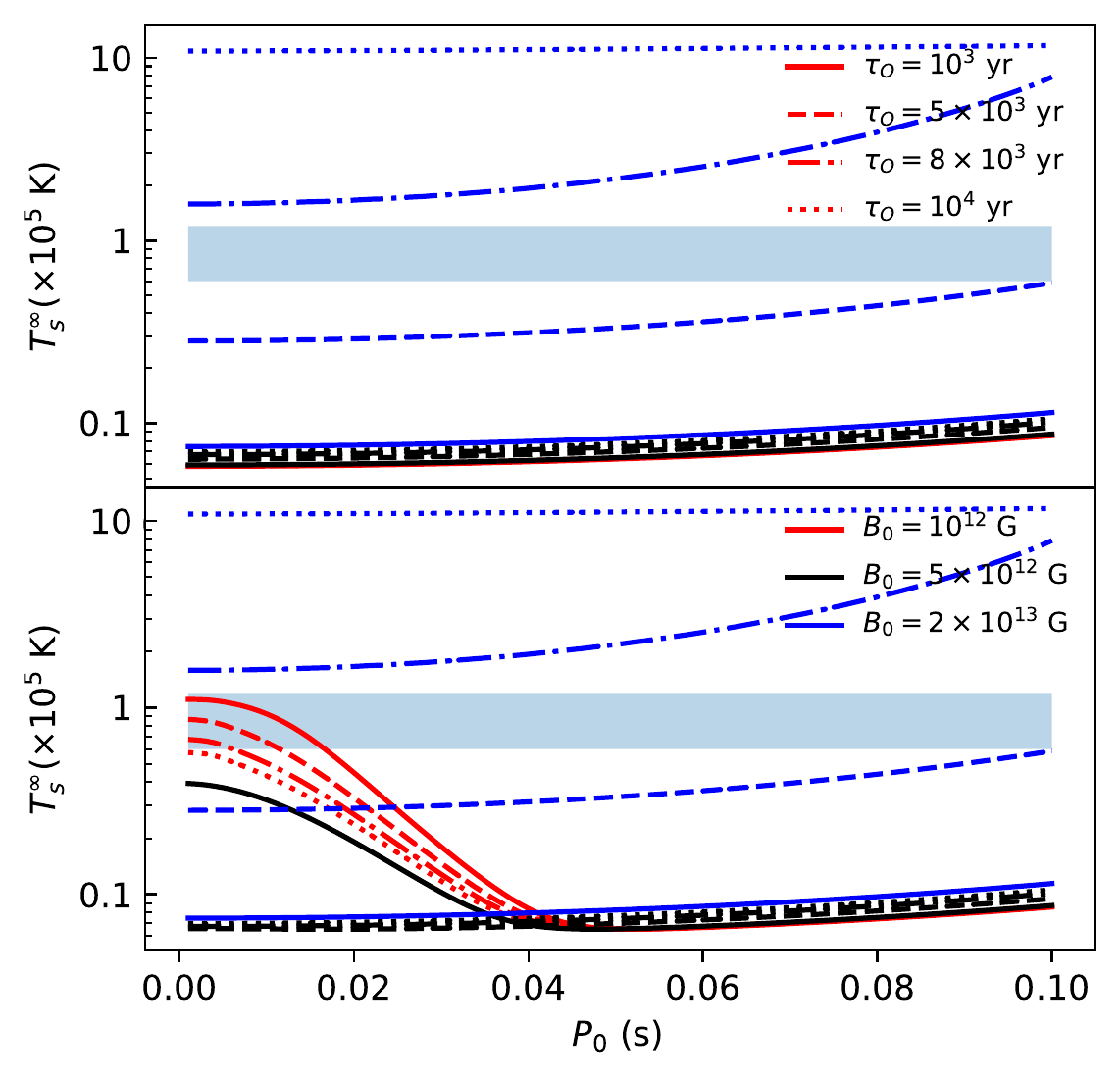}
\caption{Relations between $T_s^{\infty}$ and initial spin period of PSR B0950+08. The top panel is for the standard cooling case without any heating mechanism, and the bottom one is for the case with chemical heating. The blue shaded region is for the surface temperature proposed in \citet{Abramkin2022Apj}. The red, black and blue lines are for the different initial magnetic field $B_{\rm 0}=10^{12}\rm\ G$, $5\times10^{12}\rm\ G$ and $2\times10^{12}\rm\ G$. The solid, dashed, dash-dotted and dotted lines 
are for the different Ohmic time-scale, $\tau_{\rm O}=10^3\rm\ yr$, $5\times10^3\rm\ yr$, $8\times10^3\rm\ yr$ and $10^4\rm\ yr$.}\label{fig:Ts_exp}
\end{figure}

The thermal evolution with chemical heating will reach a quasi-steady state at the later stage in which the temperature and chemical imbalance of the stars are determined only by their current spin and spin-down rate, almost independent of their thermal evolution history \citep{1995ApJ...442..749R,Fernandez2005ApJ}. From Fig. \ref{fig:Ts_evolve_total}, it can be seen that the effect of chemical heating is still sensitive to the spin-down model and initial spin period. A stronger initial magnetic field and longer Ohmic decay time-scale will make the temperature at a quasi-steady state lower. The relations between $T_s^{\infty}$ and the initial spin period of PSR B0950+08 for different parameters of magnetic field decay model are given in Fig. \ref{fig:Ts_exp}. The model of chemical heating coupled with long-term magnetic field decay model could explain the surface temperature of PSR B0950+08 when $P_0\leq17\rm\ ms$, and the corresponding true age from the model $t_{\rm true}\sim 17.14\ \rm Myr$. 

\section{Conclusions and discussion}
\label{sec:conclusion}

\par Using the 14 years of timing observations at NSRT, we have presented the timing parameters and calculated the braking index of PSR B0950+08. The spin-frequency derivative and braking indices show a quasi-period oscillation. The braking index of PSR B0950+08 varies from --367 392 to 168 883, which shows an oscillation with large amplitude ($\sim 10^5$) and uncertainty. A model of long-term magnetic field decay modulated by a short-term oscillations is suitable for this pulsar, and three-component oscillations are more reasonable for the spin-frequency derivative distribution of PSR B0950+08. The spin evolution simulations show that the initial spin period $P_0$ must be smaller than $97\rm\ ms$ with the lower limit of kinematic age. 

\par The measurement of $n$ or $\ddot{\nu}$ of a pulsar has proved to be very difficult because various kinds of timing irregularities exist, such as glitches and a large amount of timing noise. Timing observations of 374 pulsars revealed an anomalous distribution of $\ddot{\nu}$. \citet{2020MNRAS.494.2012P} measured the braking indices of 19 young radio pulsars. Their results show a wide distribution of braking indices statistically, with a range of about $-100 \sim 3000$. As of yet, only nine young pulsars have had their braking indices measured precisely \citep{Lyne2015, Ou2016, Archibald2016ApJ}. The origins of different timing noise are still under discussion, including the short-term oscillation of the magnetic field we used in this work. 

We tested the long-term spin and thermal evolution of PSR B0950+08 with the kinematic age proposed in \citet{2019MNRAS.482.3415I}, the uncertainty of its age is also considered. The standard cooling model can explain the surface temperature of PSR B0950+08 while the age is around its most probable kinematic age. Considering relatively large characteristic age $\tau_{\rm c}$ of PSR 0950+08, two heating mechanisms are discussed. The model of vortex creep heating with a long-term magnetic field decay could maintain a relatively high temperature at the late stage of thermal evolution and explain the thermal emission data of old and warm pulsars; the effect of different spin evolution parameters on the vortex creep heating is insignificant. The effect of chemical heating mechanisms is constrained by the initial spin period. Coupling with the long-term magnetic field decay, an explanation of the temperature of PSR B0950+08 by chemical heating needs an implausibly short initial rotation period ($P_0 \leq 17\rm{ ms}$). 

As discussed in some former works \citep{Yuan1999AA, Kang2007MNRAS, Zheng2014PhLB}, deconfinement heating is also an effective heating mechanism for older but warm pulsars. The heat generation of neutron stars containing mixed hadron–quark phase are calculated. The mean value of heat per baryon is around $0.1\rm MeV$, which is one order of magnitude higher than that of chemical heating \citep{Zheng2014PhLB}. Further studies of interior structure and equation of state of neutron star coupled with thermal evolution could help us to gain more information about the thermal emission from pulsars.

The most significant uncertainty of the thermal evolution of neutron stars comes from their age. The true age of a pulsar can be different from its characteristic age. The small value of the most probable kinematic age of PSR B0950+08 proposed in \citet{2019MNRAS.482.3415I} is caused mainly by the small transverse velocity, which requires a large radial velocity at the assumed probability distribution of total velocities of pulsars. The possibility that the actual total and radial velocities of PSR B0950+08 are substantially smaller than the most probable values cannot be excluded. Measuring 3D velocities and distance of isolated radio pulsars is essential and hard to achieve. The unprecedented sensitivity of the observations with the Five-hundred-meter Aperture Spherical radio Telescope seems to provide a new opportunity.

In conclusion, PSR B0950+08 is a very peculiar source among other old classical radio pulsars.
The spin and thermal evolution are interrelated and interact with each other. The properties of its magnetic field and initial spin period are the critical factors. Thus, the spin and thermal evolution should be considered simultaneously. High-precision and long timing observations would help us to know more about the braking mechanisms and magnetospheres of pulsars. A large sample of classical pulsars with timing, ultraviolet or X-ray observations is essential for studying the evolution and interior properties of pulsars.

\section*{Acknowledgements}
We would like to thank the anonymous referee for helpful suggestions that led to significant improvement of our study. This work is supported in part by the Opening Foundation of Xinjiang Key Laboratory (No. 2021D04016), the CAS ``Light of West China'' Program (No. 2019-XBQNXZ-B-016), the National Natural Science Foundation of China (grant Nos. 12033001, 11873040, 11873080). The Nanshan 26-m Radio Telescope is partly supported by the Operation, Maintenance and Upgrading Fund for Astronomical Telescopes and Facility Instruments, budgeted from the Ministry of Finance of China (MOF) and administrated by the Chinese Academy of Sciences (CAS).

\section*{Data Availability}
The data underlying this article will be shared on reasonable
request to the corresponding author.

\bibliographystyle{mnras}
\bibliography{ref} 

\bsp	
\label{lastpage}
\end{document}